\documentclass[12pt]{article}
\usepackage{amssymb}
\usepackage{amsmath,bm}
\usepackage{graphics,mathrsfs}
\usepackage{graphicx,epsfig}
\usepackage{subfigure}
\usepackage{placeins}
\usepackage{indentfirst}
\usepackage{setspace}
\usepackage{enumerate}
\usepackage{enumitem}
\usepackage{caption}
\usepackage{varioref}
\usepackage{booktabs,tabularx}
\usepackage{color}
\usepackage{natbib}
\usepackage{epstopdf}
\usepackage{amsthm}
\makeatletter \oddsidemargin  -.1in \evensidemargin -.1in
\begin{document}
	\newcommand{\bea}{\begin{eqnarray}}
		\newcommand{\eea}{\end{eqnarray}}
	\newcommand{\nn}{\nonumber}
	\newcommand{\bee}{\begin{eqnarray*}}
		\newcommand{\eee}{\end{eqnarray*}}
	\newcommand{\lb}{\label}
	\newcommand{\nii}{\noindent}
	\newcommand{\ii}{\indent}
	\newtheorem{thm}{Theorem}[section]
	\newtheorem{example}{Example}[section]
	\newtheorem{cor}{Corollary}[section]
	\newtheorem{definition}{Definition}[section]
	\newtheorem{lemma}{Lemma}[section]
	\newtheorem{rem}{Remark}[section]
	\newtheorem{proposition}{Proposition}[section]
	\numberwithin{equation}{section}
	\renewcommand{\theequation}{\thesection.\arabic{equation}}
	\renewcommand\bibfont{\fontsize{10}{12}\selectfont}
	\setlength{\bibsep}{0.0pt}
	
	\title{\bf Reliability analysis of K-out-of-N system for Weibull components based on generalized progressive hybrid censored data$^{*}$ }
	\author{ Subhankar {\bf Dutta}\thanks {Email address :(corresponding author) subhankar.dta@gmail.com} ~and  Suchandan {\bf  Kayal}\thanks {Email address : 
			kayals@nitrkl.ac.in,~suchandan.kayal@gmail.com}
		\\{\it \small Department of Mathematics, National Institute of
			Technology Rourkela, Rourkela-769008, India}
		\\{\it \small This version has been accepeted for ESREL 2022}}
	\date{}
	\maketitle
	
	\begin{center}
		\textbf{Abstract}
	\end{center}
		In this paper, we have investigated the reliability of a K-out-of-N system for the components following Weibull distribution based on the generalized progressive hybrid censored data. We have obtained  the maximum likelihood estimates (MLEs) of the unknown parameters and the reliability function of the system. Using asymptotic normality property of MLEs, the corresponding asymptotic confidence intervals are constructed. Furthermore, Bayes estimates are derived under
		squared error loss function with informative prior by using Markov Chain Monte Carlo (MCMC) technique. Highest posterior density (HPD) credible intervals are obtained. A Monte Carlo simulation study is carried out to compare performance of the established estimates. Finally, a real data set is considered for illustrative purposes.
	\section{Introduction}\label{sec1}
	In reliability analysis and engineering practice, $K$-out-of-$N$ system is one of the most commonly used systems. So, the reliability analysis for such system is so much impactful due to it's importance in wide range of applicability. Usually two types of such systems are there. One of this is called $K$-out-of-$N: F$ system in which the system containing $N$ components fails if and only if at least $K$ ($1 \leq K \leq N$) components fail. The other one is called $K$-out-of-$N: G$ system in which the system consisting $N$ components working if and only if at least $K$ components are operational. The series ($N$-out-of-$N: G$) and the parallel ($1$-out-of-$N: G$) systems are special cases of $K$-out-of-$N: G$ system. The $K$-out-of-$N$ system has been widely used
	in both industrial and military fields. Communication system with transmitters, the multiengine plane and the wire cables made of twisted wires are real life models based on $K$-out-of-$N$  system. In the recent years, $K$-out-of-$N$ system has gained a lot of attention by many researchers. \citet{qihong2014parameter} proposed an expectation maximization algorithm to obtain the estimates of a $K$-out-of-$N$ system having  exponential components. \cite{wang2016conditional} considered a $K$-out-of-$N$ system with a cold standby component with general condition having Weibull components. The mean residual life function and survival function for such system are also derived. Zhao et al. \cite{zhao2019estimation} proposed a method to estimate the system life and the residual life of a system. \cite{roy2021reliability} discussed reliability properties of a $K$-out-of-$N$ system equipped with two cold standby components. \cite{chen2005reliability} discussed two stage weighted $K$-out-of-$N$ model with common component. \cite{franko2017reliability} discussed the reliability of a weighted $K$-out-of-$N$ system consisting two different types of components with a cold standby component. Similar more studies based on $K$-out-of-$N$ can be found in the literature, see  \cite{eryilmaz2013reliability}, \cite{cui2005generalized} and  \cite{zhao2010reliability}. \\\\
	Weibull distribution is commonly used model in life-testing analysis due to flexibility in the shape of the hazard rate function and various shapes of the probability density function (PDF).  \cite{li2008reliability} discussed the reliability estimation based on the operational data of manufacturing systems as an applicability of Weibull distribution. \cite{chiodo2006bayesian} proposed a new method to make the reliability assessment of aged electrical components based on a Bayesian approach applied to Weibull stress-strength probabilistic model. Using this applicability of Weibull distribution and $K$-out-of-$N$ system, we have considered the reliability estimation of $K$-out-of-$N$ system with Weibull component. Let, $X$ be a random variable following Weibull distribution. Then the cumulative distribution function (CDF) and PDF can be expressed respectively, as \\
	\begin{equation}
		F(x;\alpha,\beta)= 1-e^{-\beta x^{\alpha}}, ~~x>0,\alpha,\beta>0,\label{1.1}
	\end{equation} 
	and \\
	\begin{align}
		f(x;\alpha,\beta)= \alpha \beta x^{\alpha-1}e^{-\beta x^{\alpha}},~~x>0,\alpha,\beta>0. \label{1.2}
	\end{align}
	The reliability and hazard rate functions of $X$ can be written respectively, as 
	\begin{align}
		R(x;\alpha,\beta)= e^{-\beta x^{\alpha}}, ~~x>0,\alpha,\beta>0,\label{1.3}
	\end{align} 
	and 
	\begin{align}
		H(x;\alpha,\beta)= \alpha \beta x^{\alpha-1},~~x>0,\alpha,\beta>0. \label{1.4}
	\end{align}
	In life testing analysis, censoring is a common phenomenon. In practice, the most commonly used schemes are Type-I and Type-II censoring schemes. The mixture of these two censoring schemes was introduced by \cite{epstein1954truncated}, known as Type-I hybrid censoring scheme. Then the Type-II hybrid censoring scheme was introduced by \cite{childs2003exact}. In recent years hybrid censoring schemes have received considerable attention, see for example  \cite{balakrishnan2013hybrid}. Note that in these above discussed censoring schemes, experimental units
	can not be removed except the terminal points of the experiments. To avoid such drawbacks,  \cite{cohen1963progressively} introduced progressive Type-II censoring scheme. In recent years, extensive works based on progressive Type-II censoring scheme have been done, see for example \cite{balakrishnan2014art}. One major drawback of this progressive Type-II censoring scheme is that if the units are highly reliable then the experiment may take a longer time to continue. To overcome such disadvantages, progressive Type-II hybrid censoring scheme was introduced by  \cite{kundu2006analysis}. In this censoring scheme, $n$ number of units are placed in an experiment with a prefixed  time $T$, a prefixed number $m$ and a progressive censoring scheme $R_1,\cdots,R_m$ such that $\sum_{i=1}^{m}R_i=n-m$. Then the experiment will be terminated at a time $min \{T,X_{m:m:n}\}$, where $X_{m:m:n}$ denotes the time of $m$-th failure unit. In past few years, progressive hybrid censoring scheme has  gained a lot of attention by many researchers. For example one can see  \cite{lin2012inference} , \cite{hemmati2013statistical}, \cite{tomer2015estimation} and \cite{dutta2022estimation}.\\\\
	Note that in progressive Type-II hybrid censoring scheme, the experimenter may observe few failures or in a worst case scenario no failure at all. For which the efficiency of estimating parameters will be low. To overcome such drawbacks, \cite{cho2015exact} introduced generalized progressive hybrid censoring scheme (GPHCS). In this censoring scheme, $n$ number of units are placed in an experiment with a prefixed  time $T$, two prefixed numbers $k$ and $m$ ($1 \leq k <m \leq n$) and a progressive censoring scheme $R_1,\cdots,R_m$ such that $\sum_{i=1}^{m}R_i=n-m$. Then the experiment will be terminated at a time $T^{*}$ = $max\{X_{k:m:n},min \{T,X_{m:m:n}\}\}$. According to this censoring scheme, the following cases are observed.\\
	\textbf{Case I:} $\{X_{1:m:n},X_{2:m:n},\cdots,X_{m:m:n}\}$, if $X_{k:m:n}<X_{m:m:n}<T$,\\
	\textbf{Case II:} $\{X_{1:m:n},X_{2:m:n},\cdots,X_{d:m:n}\}$, if $X_{k:m:n}<T<X_{m:m:n}$,\\
	\textbf{Case III:} $\{X_{1:m:n},X_{2:m:n},\cdots,X_{k:m:n}\}$, if $T<X_{k:m:n}<X_{m:m:n}$.\\
	This scheme ensures us that at least $k$ number of failures will be observed which improves the efficiency of the estimation. In recent years, GPHCS has gained a lot of attention by the researchers. For example one can see, \cite{gorny2016exact},  \cite{kotb2018bayesian}, \cite{elshahhat2017parameters} and \cite{maswadah2021improved}.  \\\\
	In literature, one can find many researches on the reliability analysis of such system. To the best of our knowledge, the reliability analysis of this system with Weibull components under GPHCS has not been studied yet. In this paper we have considered the reliability estimation of $K$-out-of-$N$: $G$ system under generalized progressive hybrid censoring with Weibull components. The main aim of this paper is two folded. First based on GPHCS, the point estimates in view of classical and Bayesian phenomenon for unknown parameters and reliability function of the system are obtained. To obtain these point estimates we have used Newton-Raphson (NR) and Markov Chain Monte Carlo (MCMC) methods. Further using asymptotic normality properties of maximum likelihood estimates (MLEs) the asymptotic confidence intervals (ACIs) and the Bayesian highest posterior density (HPD) credible intervals are constructed. Then secondly we have carried out a Monte Carlo simulation to compare the performance of the proposed estimates. One real data set has been analyzed to illustrate the proposed methods.   \\\\ 
	The rest of the paper is organized as follows. In Section $2$,  GPHCS and the reliability function of the $K$-out-of-$N$: $G$ system have been discussed. Then the MLEs and the associated asymptotic confidence intervals of the unknown parameters and the reliability function of the system are obtained in Section $3$. In Section $4$, the Bayesian estimates and associated HPD credible intervals have been derived. Then a Monte Carlo simulation has been studied to compare the performance of the proposed estimate methods in Section $5$. In Section $6$, one real data set has been analyzed to illustrate the purposes. Some conclusions have been made in Section $7$.\\
	\section{ Reliability of $K$-out-of-$N$: $G$ system and maximum likelihood estimation} 
	In this section, the reliability function of $K$-out-of-$N$: $G$ system and the maximum likelihood estimates for the unknown parameters and reliability function based on GPHCS have been derived.  Let us consider a $K$-out-of-$N$: $G$ system consisting $N$ independent and identically distributed components. This system works if and only if at least $K$ components are operational. Let, $X_i$, for $i=1,2,\cdots,N$, denote the lifetime of the components. Here $X_i$ follows Weibull distribution with CDF and PDF as given in $(\ref{1.1})$ and $(\ref{1.2})$, respectively. It has been assumed that at initial moment all components are new and start to work at the same time. Then the reliability function of the $K$-out-of-$N$: $G$ system is given as 
	\begin{align}
		\nonumber &R_{NK}(t)\\
		\nonumber= & \sum_{i=K}^{N} \binom{N}{i} P\{X_{i+1},\cdots,X_{n}\leq t < X_1,\cdots,X_i\} \\
		\nonumber= &  \sum_{i=K}^{N} \binom{N}{i} \big[1-F(t;\alpha,\beta)\big]^{i} \big[F(t;\alpha,\beta)\big]^{N-i} \\
		= &  \sum_{i=K}^{N} \binom{N}{i} e^{-i\beta t^{\alpha}} \big[1-e^{-\beta t^{\alpha}}\big]^{N-i} . \label{2.1}
	\end{align}   
	Suppose $n$ identical units are placed on a life-testing experiment. Then the likelihood function based on GPHCS combining three different cases mentioed in the previous section can be written as 
	\begin{align}
		 L(\underline{x};\alpha,\beta) \propto~ \prod_{i=1}^{D} f(x_{i};\alpha,\beta) \big[1-F(x_{i};\alpha,\beta)\big]^{R_{i}} \big[1-F(T^*;\alpha,\beta)\big]^{R^*}, \label{2.2}
	\end{align}
	where $x_i$ denotes the time of $i$-th failure unit $x_{i:m:n}$, $\underline{x}= (x_1,x_2,\cdots,x_D)$,
	\begin{align}
		\nonumber	D=\begin{cases}
			m,~ \mbox{ Case I},\\
			d,~\mbox{ Cases II},\\
			k,~ \mbox{ Case III},
		\end{cases}
		~~~~~~\nonumber T^{*}=\begin{cases}
			X_{m:m:n}, \mbox{ Case I},\\
			T,~~~~~~~~\mbox{ Cases II},\\
			X_{k:m:n}, ~\mbox{ Case III},
		\end{cases}
	\end{align}
	and $R^{*}=n-D-\sum_{i=1}^{D} R_i$. Replacing the $f(x_{i};\alpha,\beta)$ and $F(x_{i};\alpha,\beta)$ from  $(\ref{1.1})$ and $(\ref{1.2})$ in  $(\ref{2.2})$, the likelihood function can be written as 
	\begin{align}
	 L(\underline{x};\alpha,\beta) \propto~ \alpha^D \beta^D \prod_{i=1}^{D} x_i^{\alpha-1} e^{-(R_i+1)\beta x_i^{\alpha}} e^{-\beta R^{*} T^{*\alpha}}. \label{2.3}
	\end{align} 
	The corresponding log-likelihood function can be expressed as 
	\begin{align}
		 l=\log L (\underline{x};\alpha,\beta) \propto D\log \alpha + D\log \beta + (\alpha-1) \sum_{i=1}^{D} \log x_i -\beta \sum_{i=1}^{D} (R_i+1) x_i^{\alpha} - \beta R^{*} T^{*\alpha}. \label{2.4}  
	\end{align}
	The likelihood equations are given by 
	\begin{align}
		 \frac{\partial l}{\partial \alpha}=& \frac{D}{\alpha} + \sum_{i=1}^{D} \log x_i -\beta \sum_{i=1}^{D} (R_i+1)x_i^{\alpha} \log x_i -\beta R^{*} T^{*\alpha} \log T^{*}=0, \label{2.5} \\
		\nonumber \mbox{and}~~~~~~~~~~~~~&\\
		\frac{\partial l}{\partial \beta}=&~ \frac{D}{\beta} - \sum_{i=1}^{D} (R_i+1)x_i^{\alpha}-R^{*} T^{*\alpha}=0. \label{2.6} 
	\end{align}
	These non-linear equations in $(\ref{2.5})$ and $(\ref{2.6})$ can not be solved explicitly. Thus we have to use a numerical method such as Newton-Raphson (N-R) method to obtain the MLEs of the unknown parameters as $\widehat{\alpha}$ and $\widehat{\beta}$. The MLE of the reliability function of $K$-out-of-$N$: $G$ system can be obtained by using the invariant property of MLE and is given by 
	\begin{align}
		\widehat{R}_{NK}(t)= \sum_{i=K}^{N} \binom{N}{i} e^{-i\widehat{\beta} t^{\widehat{\alpha}}} \big[1-e^{-\widehat{\beta} t^{\widehat{\alpha}}}\big]^{N-i}. \label{2.7}
	\end{align}
	
	\section{Asymptotic confidence interval}
	In this section, using asymptotic normality properties of the MLEs of unknown parameters $\alpha$ and $\beta$, the $100(1-\gamma)\%$ asymptotic confidence intervals for the unknown parameters as well as the reliability function of $K$-out-of-$N$: $G$ system have been constructed. Under some mild regularity conditions, the asymptotic distribution of the MLEs $(\widehat{\alpha},\widehat{\beta})$ can be obtained as 
	\begin{align}
		\nonumber (\widehat{\alpha},\widehat{\beta}) \sim N ((\widehat{\alpha},\widehat{\beta}), I^{-1}(\widehat{\alpha},\widehat{\beta})),
	\end{align}  
	where 
	\begin{align}
		\nonumber \widehat{I}^{-1}(\widehat{\alpha},\widehat{\beta})= {\begin{bmatrix}
				-l_{11} & -l_{12} \\
				-l_{21} & -l_{22} \\
		\end{bmatrix}}^{-1}_{(\alpha,\beta)=(\widehat{\alpha},\widehat{\beta})}\\= {\begin{bmatrix}
				Var(\widehat{\alpha}) & Cov(\widehat{\alpha},\widehat{\beta}) \\
				Cov(\widehat{\alpha},\widehat{\beta})&  	Var(\widehat{\beta})
		\end{bmatrix}},\label{eq 2.8}
	\end{align}
	is the inverse of the observed Fisher information matrix, with 
	\begin{align}
		\nonumber l_{11}=& \frac{\partial^2 l}{\partial \alpha^2} = -\frac{D}{\alpha^2}- \beta \sum_{i=1}^{D} (R_i+1)x_i^{\alpha} (\log x_i)^2\\
		&\nonumber -\beta R^{*} T^{*\alpha} (\log T^{*})^2,\\
		\nonumber l_{12}=& l_{21}=\frac{\partial^2 l}{\partial \alpha \beta} = - \sum_{i=1}^{D} (R_i+1)x_i^{\alpha} \log x_i\\
		&\nonumber - R^{*} T^{*\alpha} \log T^{*},\\
		\nonumber \mbox{and}~~~~~~~~~~&\\
		\nonumber l_{22}=& \frac{\partial^2 l}{\partial \beta^2} = -\frac{D}{\beta^2}.  
	\end{align}
	Thus, the $100(1-\gamma) \%$ asymptotic confidence intervals for $\alpha$ and $\beta$ are given by
	\begin{align}
		\nonumber \bigg(\widehat{\alpha}\underline{+}~z_{\frac{\gamma}{2}} \sqrt{Var(\widehat{\alpha})} \bigg)~~~~ \mbox{and}~~~ \nonumber \bigg(\widehat{\beta}\underline{+}~z_{\frac{\gamma}{2}} \sqrt{Var(\widehat{\beta})}\bigg),
	\end{align}
	where $z_{\frac{\gamma}{2}}$ denotes the upper $\frac{\gamma}{2}$-th percentile point of a standard normal distribution. To obtain the asymptotic confidence interval for the reliability function of $K$-out-of-$N$: $G$ system the delta method has been employed.\\
	Let us consider, $\Phi^{T}= \bigg(\frac{\partial R_{NK}(t)}{\partial \alpha},\frac{\partial R_{NK}(t)}{\partial \beta}\bigg)$, where
	\begin{align}
		\nonumber \frac{\partial R_{NK}(t)}{\partial \alpha}=&  ~ \beta t^{\alpha} \log t \sum_{i=K}^{N} \binom{N}{i} e^{-i\beta t^{\alpha}} \\
		\times& \big(1-e^{-\beta t^{\alpha}}\big)^{N-i-1} (N-2i+i e^{-\beta t^{\alpha}}),\label{3.2} 
	\end{align} 
	and 
	\begin{align}
		\nonumber \frac{\partial R_{NK}(t)}{\partial \beta}= &~ t^{\alpha} \sum_{i=K}^{N} \binom{N}{i} e^{-i\beta t^{\alpha}} \big(1-e^{-\beta t^{\alpha}}\big)^{N-i-1}\\ &\times (N-2i+i e^{-\beta t^{\alpha}}). \label{3.3}
	\end{align}  
	
	Then using delta method, the variance of $R_{NK}(t)$ can be approximated by 
	\begin{align}
		\widehat{Var}(\widehat{R}_{NK}(t)) = \bigg[\Phi^T I^{-1}(\widehat{\alpha},\widehat{\beta}) \Phi\bigg]_{(\alpha,\beta)=(\widehat{\alpha},\widehat{\beta})}. \label{3.4}
	\end{align}
	Further, $\frac{\widehat{R}_{NK}(t)-{R}_{NK}(t)}{\sqrt{\widehat{Var}(\widehat{R}_{NK}(t))}}$ follows standard normal distribution asymptotically. Thus, the $100(1-\gamma)\%$ asymptotic confidence interval for the reliability of $K$-out-of-$N$: $G$ system is given by 
	\begin{align}
		\nonumber \bigg(\widehat{R}_{NK}(t)\underline{+}~z_{\frac{\gamma}{2}} \sqrt{\widehat{Var}(\widehat{R}_{NK}(t))}\bigg), \label{3.5}
	\end{align} 
	where $z_{\frac{\gamma}{2}}$ denotes the upper $\frac{\gamma}{2}$-th percentile point of a standard normal distribution.
	
	\section{Bayesian estimation and associated credible intervals}	 
	In this section, Bayes estimates and the associated HPD credible intervals for the unknown parameters and the reliability function of the $K$-out-of-$N$: $G$ system under GPHCS have been considered. 
	\subsection{Bayes estimates}
	In this subsection, Bayes estimates have been derived based on the squared error loss function (SELF), which can be defined as 
	\begin{align}
		L(\theta,\widehat{\theta})= (\theta-\widehat{\theta})^2,
	\end{align}
	where $\widehat{\theta}$ is the estimate of $\theta$. Note that it is difficult to get joint conjugate prior distribution for the present estimation problem. Since  all the elements of the corresponding expected Fisher information matrix are not of closed forms, Jeffrey's prior can not be defined. It should be noted here that there is no clear methodology to choose an appropriate prior for Bayesian analysis. The gamma distribution is versatile for adjusting different shapes of the density function. Because of this characteristic gamma priors have been considered here as 
	\begin{align}
		\pi_1(\alpha) \propto&~ \alpha^{a-1} e^{-b\alpha}, ~~\alpha>0, a,b>0,\label{4.2}\\
		\nonumber \mbox{and}~~~~~~~~~~~~~~~~~& \\
		\pi_2(\beta) \propto&~ \beta^{c-1} e^{-d\beta}, ~~\beta>0, c,d>0. \label{4.3}
	\end{align} 
	Thus, using $(\ref{4.2})$ and $(\ref{4.3})$ the joint prior distribution of $\alpha$ and $\beta$ can be written as 
	\begin{align}
		\nonumber \pi^{*}(\alpha,\beta) \propto~& \alpha^{a-1} \beta^{c-1} e^{-b\alpha-d\beta}, ~ \alpha,\beta>0, \\
		& a,b,c,d >0. \label{4.4}
	\end{align} 
	Combining $(\ref{2.3})$ and $(\ref{4.4})$, the joint posterior distribution of $\alpha$ and $\beta$ can be expressed as 
	\begin{align}
		\nonumber \pi(\alpha,\beta|\underline{x}) &=~ \frac{\pi^{*}(\alpha,\beta) L(\underline{x};\alpha,\beta)}{\int_{0}^{\infty}\int_{0}^{\infty}\pi^{*}(\alpha,\beta) L(\underline{x};\alpha,\beta) ~d\alpha d\beta} \\
	    &= ~ K^{*-1}\alpha^{D+a-1} \beta^{D+c-1} e^{-b\alpha-d\beta} \prod_{i=1}^{D} x_i^{\alpha-1} e^{-(R_i+1)\beta x_i^{\alpha}} e^{-\beta R^{*} T^{*\alpha}}, \label{4.5}  
	\end{align}
	where, $K^{*}=\int_{0}^{\infty}\int_{0}^{\infty} \alpha^{D+a-1} \beta^{D+c-1} e^{-b\alpha-d\beta}\\
	\times \prod_{i=1}^{D} x_i^{\alpha-1} e^{-(R_i+1)\beta x_i^{\alpha}} e^{-\beta R^{*} T^{*\alpha}}~d\alpha d\beta$.
	Thus, the Bayes estimate of any function of $\alpha$ and $\beta$, say $\psi(\alpha, \beta)$ under SELF is as follows
	\begin{align}
		\widehat{\psi}_{SE}= \frac{\int_{0}^{\infty}\int_{0}^{\infty} \psi(\alpha, \beta) \pi^{*}(\alpha,\beta) L(\underline{x};\alpha,\beta) ~d\alpha d\beta}{\int_{0}^{\infty}\int_{0}^{\infty}\pi^{*}(\alpha,\beta) L(\underline{x};\alpha,\beta) ~d\alpha d\beta}. \label{4.6}
	\end{align} 
	From $(\ref{4.6})$, it is clear that the Bayes estimate of $\psi(\alpha,\beta)$ with respect to the SELF can not be obtained explicitly. Thus, an approximation technique is required to compute the desired Bayes estimates.
	
	\subsection{MCMC method} 
	From $(\ref{4.6})$, it has been observed that the Bayes estimate of $\psi(\alpha,\beta)$ can be expressed as a ratio of two integrals. It can not be solved explicitly. So, MCMC method has been employed to obtain the approximate Bayes estimates and the corresponding HPD credible intervals of the unknown parameters $\alpha$, $\beta$ and the reliability of the $K$-out-of-$N$: $G$ system. From the joint posterior density function given by $(\ref{4.5})$, the conditional posterior densities for the unknown parameters can be obtained as 
	\begin{align}
		 \pi(\alpha|\beta,\underline{x}) \propto &~ \alpha^{D+a-1} e^{-b\alpha} \prod_{i=1}^{D} x_i^{\alpha-1} e^{-(R_i+1)\beta x_i^{\alpha}} e^{-\beta R^{*}T^{*\alpha}}, \label{4.7} \\
		\nonumber \mbox{and}~~~~~~~~~~~~~~~~~~~~& \\
	 \pi(\beta|\alpha,\underline{x}) \propto &~ \beta^{D+c-1} e^{-d\beta} \prod_{i=1}^{D} e^{-(R_i+1)\beta x_i^{\alpha}} e^{-\beta R^{*}T^{*\alpha}}. \label{4.8} 
	\end{align}  
	The above conditional posterior distributions can not be reduced into the forms of some well-known distributions. Thus, we generate the posterior samples for the model parameters by using the Metropolis-Hastings (M-H) algorithm with normal proposal distributions. The following algorithm is proposed to generate samples for the unknown model parameters, and then obtain Bayes estimates.\\\\
	\textbf{Step 1:} Set initial guesses as $\alpha_0=\widehat{\alpha}$ and $\beta_0=\widehat{\beta}$.\\
	\textbf{Step 2:} Set $j=1$.\\
	\textbf{Step 3:} Generate $\alpha_j$ and $\beta_j$, where $\alpha_j \sim N(\alpha_{j-1},Var(\widehat{\alpha}))$ and $\beta_j \sim N(\beta_{j-1},Var(\widehat{\beta}))$.\\
	\textbf{Step 4:} Compute $\Omega_{\alpha}= \frac{\pi(\alpha_j|\beta_{j-1},\underline{x})}{\pi(\alpha_{j-1}|\beta_{j-1},\underline{x})}$ and accept $\alpha_j$ with the probability $min \{1,\Omega_{\alpha}\}$. Then, compute $\Omega_{\beta}= \frac{\pi(\beta_j|\alpha_j,\underline{x}) }{\pi(\beta_{j-1}|\alpha_j,\underline{x}) }$ and and accept $\beta_j$ with the probability $min \{1,\Omega_{\beta}\}$. \\
	\textbf{Step 5:} For given time $t$, compute $ R^{(j)}_{NK}(t)= \sum_{i=K}^{N} \binom{N}{i} e^{-i\beta_j t^{\alpha_j}} \big[1-e^{-\beta_j t^{\alpha_j}}\big]^{N-i} $. \\
	\textbf{Step 6:} Set $j=j+1$. \\
	\textbf{Step 7:} Repeat steps $3-6$ upto $B$ times to obtain $(\alpha_1,\cdots,\alpha_B)$, $(\beta_1,\cdots,\beta_B)$ and $(R^{(1)}_{NK}(t),\cdots, R^{(B)}_{NK}(t))$. \\
	
	Thus, the Bayes estimates of the parameters $\alpha$ and $\beta$ and the reliability function $K$-out-of-$N$: $G$ system under SELF are given as
	\begin{align}
		\nonumber \widehat{\alpha}&_{BE}= \frac{1}{B} \sum_{j=1}^{B} \alpha_j, ~~~~\widehat{\beta}_{BE}= \frac{1}{B} \sum_{j=1}^{B} \beta_j,\\
		\nonumber &\mbox{and}~~~ \widehat{R}_{NK}(t)_{BE}= \frac{1}{B} \sum_{j=1}^{B} R^{(j)}_{NK}(t), 
	\end{align}
	respectively. Furthermore, to construct the HPD credible intervals for the model parameters $\alpha$, $\beta$ and the reliability function of $K$-out-of-$N$: $G$ system ${R}_{NK}(t)$, one
	may employ the above discussed posterior samples and follow the method proposed by \cite{chen1999monte}. According to this method, the samples are rearranged in an increasing order and these ordered samples are obtained as $(\alpha^{(1)},\cdots,\alpha^{(B)})$, $(\beta^{(1)},\cdots,\beta^{(B)})$ and $(R_{NK}(t)_{1},\cdots,R_{NK}(t)_{B})$. Then, the $100(1-\gamma)\%$ HPD credible intervals can be constructed as
	\begin{align}
		\nonumber (&\alpha^{[B\gamma/2]},\alpha^{[B(1-\gamma/2)]}),~~(\beta^{[B\gamma/2]},\beta^{[B(1-\gamma/2)]}),\\
		\nonumber& \mbox{and}~~ (R_{NK}(t)_{[B\gamma/2]},R_{NK}(t)_{[B(1-\gamma/2)]}),
	\end{align} 
	where $\gamma$ is the nominal significance level and $[\cdot]$ represents the greatest integer value function. \\
	
	\section{Simulation studies}
	In this section, a Monte Carlo simulation study has been carried out to compare the performance of the proposed estimates for the unknown parameters and the reliability function of $K$-out-of-$N$: $G$ system based on $10000$ generated GPHCS using $R$ software. Without loss of generality, we have considered a $3$-out-of-$5$ system and the true values of the parameters as $\alpha= 1.5$ and $\beta=1$ for simulation. For the prefixed time $t=0.5$, the reliability function $R_{3,5}(0.5)=0.8398$. To generate the GPHCS, we have considered different values of $n,m$ and $k$ with three following progressive censoring schemes:\\
	\textbf{Scheme I:} $R_1=(n-m),R_{2}=\cdots=R_{m}=0$. \\
	\textbf{Scheme II:} $R_1=\cdots=R_{m-1}=0, R_{m}=(n-m)$.\\
	\textbf{Scheme III:} $R_{1}=\cdots=R_{n-m}=1, R_{n-m+1}=\cdots=R_{m}=0$. \\
	The performance of the estimates have been compared based on the following perspectives:
	\begin{itemize}
		\item \textbf{Mean squared error (MSE)}: $\frac{1}{N} \sum_{i=1}^{N}\big(\Theta_i-\widehat{\Theta}_i\big)^2$. The smaller value of MSE indicates the better performance of the estimates.
		\item \textbf{Average width (AW)}: Average width of the interval estimates with $\gamma$ significance level has been evaluated. Smaller width corresponds to better performance of the interval estimates.
	\end{itemize}
	In case of MLEs, to solve the non-linear likelihood equations the NR method has been used. Based on these estimates MLEs of $K$-out-of-$N$: $G$ system have been obtained. Using the normality properties of MLEs and delta method, $95\%$ confidence intervals for the parameters and the reliability function of $K$-out-of-$N$: $G$ system have been constructed. To compute the Bayes estimates, the hyper-parameters for gamma priors have been considered as $(a_1,b_1)=(3,2)$ and $(a_2,b_2)=(2.5,2.5)$, so that the prior means have become exactly equal to the true values of the parameters. To obtain the Bayes estimates under SELF, M-H algorithm has been used to generate $10000$ MCMC samples. By using these MCMC samples, $95\%$ HPD credible intervals have been constructed. Following conclusions can be drawn from Tables $\ref{T1}-\ref{T4}$: \\\\
	\textbf{(a)} For fixed $(n,m)$ and $T$, when the value of $k$ increases the values of MSEs decrease. Similarly, for the fixed values of $(n,k)$ and $T$, when the values of $m$ increases the values of MSEs decrease. While $T$ increases, for fixed $(n,m,k)$ any trend will not be observed in the values of MSE. \\
	\textbf{(b)} Bayes estimates perform better than the MLEs in terms of MSEs. \\
	\textbf{(c)} In most of the cases, MSEs of the MLEs, Bayes estimates and the reliability function under scheme I are smaller than the schemes II and III. \\
	\textbf{(d)} For fixed $(n,m)$ and $T$, when the value of $k$ increases the values of ALs decrease. Similarly, for the fixed values of $(n,k)$ and $T$, when the values of $m$ increases the values of ALs decrease. In most of the cases, while $T$ increases, for fixed $(n,m,k)$ the ALs decrease. \\
	\textbf{(e)} Under similar conditions, HPD credible intervals perform better than the ACIs in terms of AL. \\
	From these simulation study and observations we can summarize that the point and interval estimates based on Bayesian method perform better than the MLEs in terms of MSEs. Bayesian estimation contains more information than the maximum likelihood estimation.   
	
	\begin{table}
		\begin{center}
			\caption{The AEs and MSEs of estimates of $\alpha$, $\beta$ and $R_{5,3}(t)$ for different $(n, m, k)$ and $T_0=1$ under various censoring schemes (CS). }
			\label{T1}
			\tabcolsep 7pt
			\small
			\scalebox{0.75}{
				\begin{tabular}{*{13}c*{11}{r@{}l}}
					\toprule
					\multicolumn{1}{c}{$n$} & \multicolumn{1}{c}{$m$} & \multicolumn{1}{c}{$k$}
					& \multicolumn{1}{c}{CS}
					& \multicolumn{1}{c}{$\widehat{\alpha}$} & \multicolumn{1}{c}{$\widehat{\beta}$}& \multicolumn{1}{c}{$\widehat{R}_{5,3}(0.5)$} & \multicolumn{1}{c}{$\widehat{\alpha}_{BE}$}& \multicolumn{1}{c}{$\widehat{\beta}_{BE}$}& \multicolumn{1}{c}{$\widehat{R}_{5,3}(0.5)_{BE}$} \\
					\midrule
					40& 20& 10& I& 1.6214& 1.1447& 0.8220& 1.4900& 0.9924& 0.8375 \\
					& & & & (0.1501)& (0.2168)& (0.0085)& (0.0137)& (0.0126)& (0.0019) \\
					& & & II& 1.5828& 1.0552& 0.8282& 1.4915& 0.9876& 0.8390 \\
					& & & & (0.1337)& (0.1094)& (0.0117)& (0.0143)& (0.0135)& (0.0021) \\
					& & & III& 1.5850& 1.0991& 0.8226& 1.4947& 1.0003& 0.8357 \\
					& & & & (0.1238)& (0.1220)& (0.0086)& (0.0129)& (0.0146)& (0.0022)\\
					\\
					& & 15& I& 1.6483& 1.1399& 0.8285& 1.4937& 0.9887& 0.8394 \\
					& & & & (0.1486)& (0.1896)& (0.0079)& (0.0135)& (0.0123)& (0.0019) \\
					& & & II& 1.4545& 1.0886& 0.7935& 1.4944& 1.0081& 0.8328 \\
					& & & & (0.1220)& (0.0925)& (0.0112)& (0.0141)& (0.0134)& (0.0020) \\
					& & & III& 1.5837& 1.0878& 0.8258& 1.4990& 0.9862& 0.8413 \\
					& & & & (0.1153)& (0.1204)& (0.0084)& (0.0125)& (0.0130)& (0.0019) \\
					\midrule 
					& 30& 10& I& 1.5680& 1.0329& 0.8361& 1.4946& 0.9929& 0.8383 \\
					& & & & (0.0925)& (0.0511)& (0.0059)& (0.0131)& (0.0123)& (0.0018) \\
					& & & II& 1.5622& 1.0372& 0.8311& 1.4962& 0.9938& 0.8380 \\
					& & & & (0.1079)& (0.0654)& (0.0080)& (0.0134)& (0.0141)& (0.0021) \\
					& & & III& 1.5512& 1.0399& 0.8293& 1.4905& 0.9930& 0.8372 \\
					& & & & (0.1059)& (0.0675)& (0.0076)& (0.0138)& (0.0142)& (0.0020)\\
					\\
					& & 15& I& 1.5504& 1.0301& 0.8340& 1.4864& 0.9972& 0.8350 \\
					& & & & (0.0865)& (0.0500)& (0.0057)& (0.0128)& (0.0118)& (0.0018) \\
					& & & II& 1.5592& 1.0400& 0.8307& 1.4943& 0.9943& 0.8372 \\
					& & & & (0.1065)& (0.0649)& (0.0078)& (0.0130)& (0.0124)& (0.0021) \\
					& & & III& 1.5652& 1.0530& 0.8294& 1.4998& 0.9938& 0.8388 \\
					& & & & (0.1001)& (0.0639)& (0.0070)& (0.0134)& (0.0133)& (0.0020) \\
					\midrule
					80& 50& 30& I& 1.5568& 1.0310& 0.8391& 1.4998& 0.9980& 0.8383 \\
					& & & & (0.0470)& (0.0295)& (0.0029)& (0.0124)& (0.0103)& (0.0015) \\
					& & & II& 1.5182& 1.0182& 0.8331& 1.4919& 0.9953& 0.8366 \\
					& & & & (0.0525)& (0.0338)& (0.0043)& (0.0133)& (0.0118)& (0.0019) \\
					& & & III& 1.5454& 1.0414& 0.8334& 1.4981& 0.9960& 0.8385 \\
					& & & & (0.0518)& (0.0403)& (0.0036)& (0.0127)& (0.0115)& (0.0016) \\
					\\
					& & 40& I& 1.5448& 1.0255& 0.8384& 1.4979& 0.9936& 0.8393  \\
					& & & & (0.0439)& (0.0282)& (0.0029)& (0.0118)& (0.0105)& (0.0015)\\
					& & & II& 1.3646& 1.1035& 0.7650& 1.4973& 1.0402& 0.8235 \\
					& & & & (0.0590)& (0.0378)& (0.0098)& (0.0130)& (0.0125)& (0.0018) \\
					& & & III& 1.4543& 1.0480& 0.8110& 1.4998& 1.0141& 0.8329 \\
					& & & & (0.0498)& (0.0380)& (0.0043)& (0.0122)& (0.0111)& (0.0016) \\
					\midrule
					& 60& 30& I& 1.5361& 1.0068& 0.8411& 1.5027& 0.9914& 0.8410 \\
					& & & & (0.0435)& (0.0210)& (0.0029)& (0.0119)& (0.0103)& (0.0015) \\
					& & & II& 1.5312& 1.0120& 0.8372& 1.5004& 0.9931& 0.8396 \\
					& & & & (0.0497)& (0.0323)& (0.0042)& (0.0132)& (0.0118)& (0.0018) \\
					& & & III& 1.5366& 1.0178& 0.8379& 1.4969& 0.9949& 0.8384 \\
					& & & & (0.0482)& (0.0291)& (0.0035)& (0.0127)& (0.0112)& (0.0015) \\
					\\
					& & 40& I& 1.5334& 1.0141& 0.8387& 1.4968& 1.0036& 0.8355 \\
					& & & & (0.0423)& (0.0203)& (0.0028)& (0.0118)& (0.0101)& (0.0015) \\
					& & & II& 1.4683& 1.0241& 0.8217& 1.4966& 1.0074& 0.8343 \\
					& & & & (0.0493)& (0.0293)& (0.0035)& (0.0130)& (0.0101)& (0.0016) \\
					& & & III& 1.5005& 1.0338& 0.8263& 1.4976& 1.0041& 0.8358 \\
					& & & & (0.0433)& (0.0275)& (0.0034)& (0.0116)& (0.0099)& (0.0015) \\
					\bottomrule
			\end{tabular}}
		\end{center}
		\vspace{-0.5cm}
	\end{table}

	\begin{table}
		\begin{center}
			\caption{The $95\%$ ACIs, HPD credible intervals and AWs of the intervals for $\alpha$, $\beta$ and $R_{5,3}(t)$ for different $(n, m, k)$ and $T_0=1$	under various censoring schemes (CS).}
			\label{T2}
			\tabcolsep 7pt
			\small
			\scalebox{0.75}{
				\begin{tabular}{*{13}c*{11}{r@{}l}}
					\toprule
					\multicolumn{4}{c}{} &
					\multicolumn{2}{c}{$\alpha$} & \multicolumn{2}{c}{$\beta$} & \multicolumn{2}{c}{$R_{5,3}(0.5)$} \\
					\cmidrule(lr){5-6}\cmidrule(lr){7-8} \cmidrule(lr){9-10} 
					\multicolumn{1}{c}{$n$} &\multicolumn{1}{c}{$m$} & \multicolumn{1}{c}{$k$}
					& \multicolumn{1}{c}{CS} & \multicolumn{1}{c}{ACI} & \multicolumn{1}{c}{HPD}& \multicolumn{1}{c}{ACI} & \multicolumn{1}{c}{HPD}& \multicolumn{1}{c}{ ACI}& \multicolumn{1}{c}{HPD} \\
					\midrule
					40& 20& 10& I& (0.9618, 2.2810)& (1.2865, 1.7473)& (0.6589, 1.9035)& (0.7657, 1.2086)& (0.7504, 0.9180)& (0.8078, 0.8561) \\
					& & & & 1.3192& 0.4608& 1.2446& 0.4429& 0.1676& 0.0483 \\ 
					& & & II& (0.8927, 2.2730)& (1.2476, 1.7080)& (0.5270, 1.7233)& (0.7724, 1.2116)& (0.7477, 0.9221)& (0.8071, 0.8693) \\
					& & & & 1.3803& 0.4604& 1.1963& 0.4392&  0.1744& 0.0621 \\
					& & & III& (0.9601, 2.2098)& (1.2514, 1.6996)& (0.5291, 1.6873)& (0.7806, 1.2379)& (0.7348, 0.9111)& (0.8074, 0.8579)\\
					& & & & 1.2496& 0.4481& 1.1582& 0.4572& 0.1762& 0.0504 \\
					\\
					& & 15& I& (0.9779, 2.2887)& (1.2886, 1.7367)& (0.4587, 1.6923)& (0.7484, 1.1864)& (0.7553, 0.9173)& (0.8140, 0.8529) \\
					& & & & 1.3108& 0.4480& 1.2336& 0.4380& 0.1619& 0.0389 \\
					& & & II& (0.8508, 2.0582)& (1.2697, 1.7239)& (0.4315, 1.5618)& (0.7865, 1.2124)& (0.7386, 0.9090)& (0.7753, 0.8517) \\
					& & & & 1.2074& 0.4542& 1.1303& 0.4259& 0.1704& 0.0564 \\
					& & & III& (0.9587, 2.1987)& (1.2891, 1.7321)& (0.6384, 1.7837)& (0.7756, 1.2019)& (0.7517, 0.9145)& (0.8101, 0.8515) \\
					& & & & 1.2399& 0.4429& 1.1453& 0.4263& 0.1628&  0.0413 \\ 
					\midrule
					& 30& 10& I& (1.0165, 2.1195)& (1.2518, 1.6987)& (0.5335, 1.3609)& (0.7604, 1.1888)& (0.7556, 0.9176)& (0.8225, 0.8697) \\
					& & & & 1.1030& 0.4468& 0.8274& 0.4284& 0.1620& 0.0471 \\ 
					& & & II& (0.9580, 2.1664)& (1.2551, 1.7047)& (0.8813, 1.8372)& (0.7851, 1.2094)& (0.7471, 0.9213)& (0.8149, 0.8674) \\
					& & & & 1.2084& 0.4496& 0.9559& 0.4243& 0.1741& 0.0525 \\
					& & & III& (0.9715, 2.1309)& (1.2764, 1.7117)& (0.9335, 1.8978)& (0.7607, 1.2130)& (0.7566, 0.9225)& (0.8144, 0.8642)\\
					& & & & 1.1594& 0.4352& 0.9643& 0.4523& 0.1658& 0.0497 \\   
					\\
					& & 15& I& (1.0053, 2.0956)& (1.2686, 1.7091)& (0.3761, 1.2019)& (0.7824, 1.2091)& (0.7511, 0.9116)& (0.8209, 0.8572) \\
					& & & & 1.0902& 0.4405& 0.8257& 0.4266& 0.1605& 0.0363 \\
					& & & II& (0.9580, 2.1604)& (1.2693, 1.7138)& (0.4266, 1.3760)& (0.7897, 1.2084)& (0.7669, 0.9359)& (0.8147, 0.8567) \\
					& & & & 1.2024& 0.4445& 0.9494& 0.4187& 0.1690& 0.0519 \\
					& & & III& (0.9934, 2.1470)& (1.2911, 1.7249)& (0.7489, 1.7020)& (0.7629, 1.1819)& (0.7478, 0.9094)& (0.8149, 0.8439) \\
					& & & & 1.1535& 0.4338& 0.9531& 0.4189& 0.1615& 0.0389 \\
					\midrule
					80& 50& 30& I& (1.1604, 1.9532)& (1.2853, 1.7192)& (0.6218, 1.2310)& (0.8116, 1.1864)& (0.7615, 0.9079)& (0.8311, 0.8672)\\
					& & & & 0.7928& 0.4339& 0.6092& 0.3748& 0.1464& 0.0361 \\
					& & & II& (1.0713, 1.9652)& (1.2523, 1.7168)& (0.5223, 1.2506)& (0.8002, 1.2145)& (0.7513, 0.9168)& (0.8226, 0.8636) \\
					& & & & 0.8938& 0.4644& 0.7282& 0.4142& 0.1655& 0.0409 \\
					& & & III& (1.1209, 1.9699)& (1.2661, 1.7040)& (0.4724, 1.2156)& (0.7847, 1.1961)& (0.7564, 0.9098)& (0.8248, 0.8621) \\
					& & & & 0.8489& 0.4379& 0.7431& 0.4114& 0.1533& 0.0372 \\
					\\
					& & 40& I& (1.1509, 1.9387)& (1.2986, 1.7088)& (0.4910, 1.0968)& (0.7959, 1.1625)& (0.7638, 0.9053)& (0.8304, 0.8663) \\
					& & & & 0.7878& 0.4101& 0.6057& 0.3666& 0.1415&  0.0358 \\
					& & & II& (0.9904, 1.7388)& (1.2786, 1.7166)& (0.8400, 1.5565)& (0.8622, 1.2625)& (0.7375, 0.8962)& (0.7562, 0.7939) \\
					& & & & 0.7483& 0.4380& 0.7165& 0.4003& 0.1586& 0.0376 \\
					& & & III& (1.0681, 1.8404)& (1.3030, 1.7119)& (0.6969, 1.4286)& (0.8193, 1.2207)& (0.7577, 0.9089)& (0.8036, 0.8383) \\
					& & & & 0.7722& 0.4089& 0.7317& 0.4013& 0.1512& 0.0347 \\
					\midrule
					& 60& 30& I& (1.1537, 1.9185)& (1.3027, 1.7245)& (0.4994, 1.0696)& (0.7908, 1.1552)& (0.7693, 0.9075)& (0.8329, 0.8683) \\
					& & & & 0.7648& 0.4218& 0.5702& 0.3644& 0.1381& 0.0354 \\
					& & & II& (1.1060, 1.9564)& (1.2683, 1.7130)& (0.8358, 1.4976)& (0.7566, 1.1607)& (0.7569, 0.9116)& (0.8273, 0.8621) \\
					& & & & 0.8503& 0.4447& 0.6618& 0.4041& 0.1546& 0.0357 \\
					& & & III& (1.1302, 1.9431)& (1.2787, 1.7087)& (0.8109, 1.4798)& (0.8008, 1.1954)& (0.7591, 0.9108)& (0.8292, 0.8657) \\
					& & & & 0.8128& 0.4300& 0.6688& 0.3946& 0.1517& 0.0365 \\
					\\
					& & 40& I& (1.1524, 1.9144)& (1.2998, 1.7202)& (0.4930, 1.0569)& (0.8221, 1.1731)& (0.7723, 0.9083)& (0.8306, 0.8659) \\
					& & & & 0.7620& 0.4204& 0.5639& 0.3509& 0.1360& 0.0353 \\
					& & & II& (1.0684, 1.8082)& (1.2742, 1.7054)& (0.6557, 1.3209)& (0.8181, 1.1965)& (0.7593, 0.9111)& (0.8137, 0.8497) \\
					& & & & 0.7398& 0.4312& 0.6652& 0.3784& 0.1518& 0.0360 \\
					& & & III& (1.1092, 1.8919)& (1.2848, 1.7036)& (0.6541, 1.3280)& (0.8080, 1.1913)& (0.7612, 0.9109)& (0.8185, 0.8521) \\
					& & & & 0.7827& 0.4188& 0.6738& 0.3833& 0.1496& 0.0336 \\
					\bottomrule
			\end{tabular}}
		\end{center}
		\vspace{-0.5cm}
	\end{table}
	
	\begin{table}
		\begin{center}
			\caption{The AEs and MSEs of estimates of of $\alpha$, $\beta$ and $R_{5,3}(t)$ for different $(n, m, k)$ and $T_0=1.5$ under various censoring schemes (CS). }
			\label{T3}
			\tabcolsep 7pt
			\small
			\scalebox{0.75}{
				\begin{tabular}{*{13}c*{11}{r@{}l}}
					\toprule
					\multicolumn{1}{c}{$n$} & \multicolumn{1}{c}{$m$} & \multicolumn{1}{c}{$k$}
					& \multicolumn{1}{c}{CS}
					& \multicolumn{1}{c}{$\widehat{\alpha}$} & \multicolumn{1}{c}{$\widehat{\beta}$}& \multicolumn{1}{c}{$\widehat{R}_{5,3}(0.5)$} & \multicolumn{1}{c}{$\widehat{\alpha}_{BE}$}& \multicolumn{1}{c}{$\widehat{\beta}_{BE}$}& \multicolumn{1}{c}{$\widehat{R}_{5,3}(0.5)_{BE}$} \\
					\midrule
					40& 20& 10& I& 1.6467& 1.1505& 0.8234& 1.4930& 0.9940& 0.8372 \\
					& & & & (0.1696)& (0.1710)& (0.0085)& (0.0139)& (0.0122)& (0.0020) \\
					& & & II& 1.5648& 1.0353& 0.8272& 1.5168& 1.0044& 0.8390 \\
					& & & & (0.0972)& (0.0854)& (0.0128)& (0.0157)& (0.0140)& (0.0022) \\
					& & & III& 1.6037& 1.0825& 0.8290& 1.5031& 0.9979& 0.8383 \\
					& & & & (0.1165)& (0.1073)& (0.0091)& (0.0141)& (0.0133)& (0.0019) \\
					\\
					& & 15& I& 1.6503& 1.1601& 0.8250& 1.4963& 0.9948& 0.8380 \\
					& & & & (0.1651)& (0.1672)& (0.0082)& (0.0136)& (0.0118)& (0.0019) \\
					& & & II& 1.5704& 1.0559& 0.8220& 1.5226& 1.0037& 0.8407 \\
					& & & & (0.0941)& (0.0812)& (0.0118)& (0.0130)& (0.0132)& (0.0020) \\
					& & & III& 1.6029& 1.0795& 0.8294& 1.5060& 0.9919& 0.8409 \\
					& & & & (0.1121)& (0.1041)& (0.0088)& (0.0127)& (0.0130)& (0.0019) \\
					\midrule
					& 30& 10& I& 1.5820& 1.0402& 0.8364& 1.4949& 0.9919& 0.8385 \\
					& & & & (0.0704)& (0.0490)& (0.0061)& (0.0129)& (0.0111)& (0.0018) \\
					& & & II& 1.5598& 1.0136& 0.8360& 1.5430& 1.0088& 0.8437 \\
					& & & & (0.0696)& (0.0457)& (0.0073)& (0.0155)& (0.0120)& (0.0017) \\
					& & & III& 1.5640& 1.0286& 0.8341& 1.5368& 1.0069& 0.8427 \\
					& & & & (0.0717)& (0.0455)& (0.0068)& (0.0124)& (0.0130)& (0.0019) \\
					\\
					& & 15& I& 1.5830& 1.0442& 0.8345& 1.4999& 0.9912& 0.8396 \\
					& & & & (0.0646)& (0.0460)& (0.0053)& (0.0120)& (0.0110)& (0.0017) \\
					& & & II& 1.5460& 1.0129& 0.8330& 1.5330& 1.0132& 0.8397 \\
					& & & & (0.0676)& (0.0389)& (0.0072)& (0.0154)& (0.0119)& (0.0017) \\
					& & & III& 1.5510& 1.0214& 0.8335& 1.5335& 1.0083& 0.8418 \\
					& & & & (0.0690)& (0.0426)& (0.0062)& (0.0119)& (0.0125)& (0.0018) \\
					\midrule
					80& 50& 30& I& 1.5592& 1.0386& 0.8364& 1.5036& 1.0027& 0.8374 \\
					& & & & (0.0948)& (0.0281)& (0.0052)& (0.0128)& (0.0094)& (0.0015) \\
					& & & II& 1.5207& 1.0109& 0.8338& 1.5530& 1.0177& 0.8433 \\
					& & & & (0.0368)& (0.0261)& (0.0045)& (0.0151)& (0.0101)& (0.0016) \\
					& & & III& 1.5252& 1.0097& 0.8378& 1.5393& 1.0158& 0.8414 \\
					& & & & (0.0336)& (0.0242)& (0.0033)& (0.0138)& (0.0104)& (0.0014) \\
					\\
					& & 40& I& 1.2956& 0.4157& 0.9567& 1.5344& 0.7140& 0.9249 \\
					& & & & (0.0914)& (0.0234)& (0.0041)& (0.0126)& (0.0085)& (0.0015) \\
					& & & II& 1.5214& 1.0175& 0.8317& 1.5515& 1.0269& 0.8400 \\
					& & & & (0.0353)& (0.0225)& (0.0043)& (0.0143)& (0.0096)& (0.0016) \\
					& & & III& 1.5289& 1.0253& 0.8332& 1.5375& 1.0158& 0.8404 \\
					& & & & (0.0315)& (0.0233)& (0.0032)& (0.0135)& (0.0103)& (0.0014) \\
					\midrule
					& 60& 30& I& 1.5384& 1.0164& 0.8383& 1.4968& 1.0003& 0.8368 \\
					& & & & (0.0334)& (0.0201)& (0.0030)& (0.0105)& (0.0085)& (0.0015) \\
					& & & II& 1.5208& 1.0194& 0.8313& 1.5626& 1.0235& 0.8439 \\
					& & & & (0.0332)& (0.0225)& (0.0043)& (0.0149)& (0.0097)& (0.0014) \\
					& & & III& 1.5245& 1.0167& 0.8343& 1.5579& 1.0197& 0.8441 \\
					& & & & (0.0309)& (0.0235)& (0.0028)& (0.0126)& (0.0094)& (0.0014) \\
					\\
					& & 40& I& 1.5391& 1.0209& 0.8369& 1.4995& 0.9961& 0.8385 \\
					& & & & (0.0301)& (0.0200)& (0.0029)& (0.0101)& (0.0083)& (0.0015) \\ 
					& & & II& 1.5249& 1.0135& 0.8346& 1.5607& 1.0245& 0.8431 \\
					& & & & (0.0328)& (0.0220)& (0.0039)& (0.0141)& (0.0095)& (0.0014) \\
					& & & III& 1.5272& 1.0142& 0.8356& 1.5618& 1.0241& 0.8435 \\
					& & & & (0.0302)& (0.0220)& (0.0026)& (0.0119)& (0.0093)& (0.0014) \\

					\bottomrule
			\end{tabular}}
		\end{center}
		\vspace{-0.5cm}
	\end{table}

	\begin{table}
		\begin{center}
			\caption{The $95\%$ ACIs, HPD credible intervals and AWs of the intervals for $\alpha$, $\beta$ and $R_{5,3}(t)$ for different $(n, m, k)$ and $T_0=1.5$	under various censoring schemes (CS).}
			\label{T4}
			\tabcolsep 7pt
			\small
			\scalebox{0.75}{
				\begin{tabular}{*{13}c*{11}{r@{}l}}
					\toprule
					\multicolumn{4}{c}{} &
					\multicolumn{2}{c}{$\alpha$} & \multicolumn{2}{c}{$\beta$} & \multicolumn{2}{c}{$R_{5,3}(0.5)$} \\
					\cmidrule(lr){5-6}\cmidrule(lr){7-8} \cmidrule(lr){9-10} 
					\multicolumn{1}{c}{$n$} &\multicolumn{1}{c}{$m$} & \multicolumn{1}{c}{$k$}
					& \multicolumn{1}{c}{CS} & \multicolumn{1}{c}{ACI} & \multicolumn{1}{c}{HPD}& \multicolumn{1}{c}{ACI} & \multicolumn{1}{c}{HPD}& \multicolumn{1}{c}{ ACI}& \multicolumn{1}{c}{HPD} \\
					\midrule
					40& 20& 10& I& (0.9780, 2.2854)& (1.2669, 1.7249)& (0.4972, 1.7407)& (0.7796, 1.2102)& (0.7540, 0.9199)& (0.8089, 0.8578) \\
					& & & & 1.3074& 0.4580& 1.2435& 0.4306& 0.1659& 0.0389 \\
					& & & II& (0.9930, 2.1366)& (1.2448, 1.7264)& (0.3991, 1.5879)& (0.7821, 1.2538)& (0.7495, 0.9283)& (0.8068, 0.8496) \\
					& & & & 1.1435& 0.4815& 1.1887& 0.4716& 0.1788& 0.0428 \\
					& & & III& (1.0356, 2.1819)& (1.2845, 1.7428)& (0.8114, 1.8375)& (0.7885, 1.2371)& (0.7529, 0.9238)& (0.8135, 0.8465) \\
					& & & & 1.1463& 0.4583& 1.0260& 0.4486& 0.1709& 0.0329 \\
					\\
					& & 15& I& (0.9800, 2.2807)& (1.2730, 1.7205)& (0.4874, 1.7223)& (0.7821, 1.2071)& (0.7595, 0.9239)& (0.8108, 0.8391) \\
					& & & & 1.3007& 0.4475& 1.2348& 0.4250& 0.1643& 0.0282 \\
					& & & II& (1.0010, 2.1397)& (1.3084, 1.7380)& (0.5036, 1.5117)& (0.8011, 1.2411)& (0.7488, 0.9209)& (0.8014, 0.8425) \\
					& & & & 1.1386& 0.4296& 1.0080& 0.4400& 0.1721& 0.0411 \\
					& & & III& (1.0335, 2.1723)& (1.2669, 1.7103)& (0.6969, 1.7221)& (0.7721, 1.2170)& (0.7614, 0.9266)& (0.8133, 0.8455) \\
					& & & & 1.1387& 0.4434& 1.0251& 0.4448& 0.1652& 0.0321 \\
					\midrule
					& 30& 10& I& (1.0857, 2.0842)& (1.2849, 1.7365)& (1.1463, 1.9257)& (0.7942, 1.2205)& (0.7520, 0.9133)& (0.8227, 0.8501) \\
					& & & & 0.9985& 0.4516& 0.7794& 0.4262& 0.1613& 0.0273 \\
					& & & II& (1.0632, 2.0565)& (1.3067, 1.7663)& (0.6876, 1.4815)& (0.7930, 1.2351)& (0.7644, 0.9217)& (0.8201, 0.8520) \\
					& & & & 0.9932& 0.4596& 0.7939& 0.4421& 0.1573& 0.0319 \\
					& & & III& (1.0848, 2.0431)& (1.2947, 1.7444)& (0.2634, 1.0612)& (0.7820, 1.2064)& (0.7631, 0.9277)& (0.8195, 0.8496) \\
					& & & & 0.9583& 0.4497& 0.7978& 0.4243& 0.1645& 0.0301 \\
					\\
					& & 15& I& (1.0864, 2.0797)& (1.2705, 1.7184)& (0.7432, 1.5048)& (0.7764, 1.1885)& (0.7534, 0.9119)& (0.8208, 0.8481) \\
					& & & & 0.9933& 0.4478& 0.7615& 0.4120& 0.1585& 0.0273 \\
					& & & II& (1.0534, 2.0386)& (1.3100, 1.7620)& (0.5991, 1.3927)& (0.8050, 1.2387)& (0.7525, 0.9027)& (0.8173, 0.8486) \\
					& & & & 0.9851& 0.4519& 0.7936& 0.4336& 0.1502& 0.0312 \\
					& & & III& (1.0740, 2.0280)& (1.3110, 1.7541)& (0.3918, 1.1852)& (0.8060, 1.2203)& (0.7621, 0.9255)& (0.8187, 0.8483) \\
					& & & & 0.9539& 0.4431& 0.7934& 0.4142& 0.1634& 0.0295 \\
					\midrule
					80& 50& 30& I& (1.1691, 1.9494)& (1.2840, 1.7262)& (0.4877, 1.0926)& (0.8232, 1.1916)& (0.7629, 0.9126)& (0.8283, 0.8544) \\
					& & & & 0.7803& 0.4422& 0.6048& 0.3684& 0.1496& 0.0261 \\
					& & & II& (1.1493, 1.8921)& (1.3575, 1.7960)& (0.7759, 1.3988)& (0.8197, 1.2181)& (0.7717, 0.9224)& (0.8233, 0.8443) \\
					& & & & 0.7428& 0.4385& 0.6229& 0.3984& 0.1507& 0.0210 \\
					& & & III& (1.1733, 1.8770)& (1.3243, 1.7517)& (0.4539, 1.0783)& (0.8344, 1.2443)& (0.7704, 0.9129)& (0.8290, 0.8465) \\
					& & & & 0.7036& 0.4274& 0.6244& 0.4098& 0.1425& 0.0175 \\
					\\
					& & 40& I& (0.9888, 1.6023)& (1.3353, 1.7481)& (0.2530, 0.6127)& (0.5640, 0.8510)& (0.8819, 0.9648)& (0.9395, 0.9639)\\
					& & & & 0.6134& 0.4127& 0.3597& 0.2869& 0.0828& 0.0244 \\
					& & & II& (1.1512, 1.8916)& (1.3298, 1.7621)& (0.6699, 1.2872)& (0.8251, 1.2134)& (0.7606, 0.9099)& (0.8217, 0.8417) \\
					& & & & 0.7403& 0.4322& 0.6172& 0.3883& 0.1492& 0.0200 \\
					& & & III& (1.1775, 1.8802)& (1.3089, 1.7231)& (0.7223, 1.3356)& (0.8178, 1.2199)& (0.7878, 0.9203)& (0.8246, 0.8518) \\
					& & & & 0.7027& 0.4141& 0.6132& 0.4020& 0.1324& 0.0272 \\ 
					\midrule
					& 60& 30& I& (1.1978, 1.8830)& (1.3088, 1.7283)& (0.6167, 1.1356)& (0.8166, 1.1691)& (0.7613, 0.9072)& (0.8303, 0.8502) \\
					& & & & 0.6852& 0.4195& 0.5188& 0.3525& 0.1459& 0.0199 \\
					& & & II& (1.1750, 1.8666)& (1.3779, 1.8065)& (0.6758, 1.2394)& (0.8277, 1.2016)& (0.7701, 0.9163)& (0.8220, 0.8406) \\
					& & & & 0.6916& 0.4286& 0.5635& 0.3738& 0.1462& 0.0186 \\
					& & & III& (1.1930, 1.8561)& (1.3533, 1.7666)& (0.7517, 1.3094)& (0.8395, 1.2053)& (0.7668, 0.9025)& (0.8256, 0.8449) \\
					& & & & 0.6630& 0.4133& 0.5577& 0.3658& 0.1357& 0.0193 \\ 
					\\
					& & 40& I& (1.1985, 1.8796)& (1.2895, 1.7027)& (0.8450, 1.3562)& (0.8193, 1.1677)& (0.7684, 0.9106)& (0.8288, 0.8450) \\
					& & & & 0.6811& 0.4131& 0.5112& 0.3483& 0.1422& 0.0162 \\ 
					& & & II& (1.1879, 1.8719)& (1.3488, 1.7699)& (0.6798, 1.2409)& (0.8476, 1.2110)& (0.7704, 0.9107)& (0.8252, 0.8440) \\
					& & & & 0.6840& 0.4211& 0.5610& 0.3633& 0.1403& 0.0188 \\
					& & & III& (1.1948, 1.8495)& (1.3924, 1.8017)& (0.5304, 1.0868)& (0.8702, 1.2263)& (0.7866, 0.9121)& (0.8269, 0.8442) \\
					& & & & 0.6547& 0.4093& 0.5564& 0.3560& 0.1255& 0.0173 \\
					\bottomrule
			\end{tabular}}
		\end{center}
		\vspace{-0.5cm}
	\end{table}
	
	\section{Real data anlysis}
	In this section, a real life data set from  \cite{linhart1986model} has been considered to illustrate the applicability of the proposed methods. This data set represents the failure times of the air conditioning system of an airplane.  The
	data set is given below: \\
	-----------------------------------------------------------------------------------------------------------------------\\
	1, 3, 5, 7, 11, 11, 11, 12, 14,
	14, 14, 16, 16, 20, 21, 23, 42, 47, 52, 62, 71, 71, 87, 90,
	95, 120, 120, 225, 246, 261. \\
	-----------------------------------------------------------------------------------------------------------------------\\
	In recent years this data set has been also analyzed by \cite{singh2015reliability}, \cite{mohamed2018inference} and \cite{ashour2020inferences}. Before performing the data analysis, we test the goodness-of-fit of this set of data with the Weibull distribution using the Kolmogorov-Smirnov (K-S) statistic. The K-S statistic and the corresponding $p$-values are obtained as 0.153 and 0.481 respectively, which show that this data set fits Weibull distribution well.\\\\
	To analyse real data set, firstly a progressive type II sample has been generated for $n=30$, $m=20$, $R_1=\cdots=R_5=2$, $R_6=\cdots=R_{20}=0$. Then the progressive type II censored sample is: 1,3, 5,
	7, 11, 14, 16, 20, 23, 42, 47, 52, 62, 71, 87, 90, 95, 120,
	225, 246. To generate GPHC sample, set $k=12, T=80$ for case-I, $k=13, T=100$ for case-II, and $k=14, T=120$ for case-III. Based on the GPHC sample generated from the real data set, 
	point estimates of $\alpha$, $\beta$ and reliability function of the
	system have been computed. To obtain Bayes estimates of $\alpha$, $\beta$ and reliability
	function of the system under SELF, non-informative priors have been considered. All of these point and interval estimates are tabulated in Tables $5$ and $6$. From Table $5$, we can observe that MLEs and Bayes estimates of the parameters] and reliability function of the system are very close. Furthermore, it has been observed that the length of HPD credible intervals are smaller than ACIs from Table $6$. 
	
	\begin{table}
		\begin{center}
			\caption{The MLEs and Bayes estimates of of $\alpha$, $\beta$ and $R_{5,3}(t)$ under various censoring schemes (CS). }
			\label{T5}
			\tabcolsep 7pt
			\small
			\scalebox{1}{
				\begin{tabular}{*{13}c*{11}{r@{}l}}
					\toprule
					\multicolumn{1}{c}{CS}
					& \multicolumn{1}{c}{$\widehat{\alpha}$} & \multicolumn{1}{c}{$\widehat{\beta}$}& \multicolumn{1}{c}{$\widehat{R}_{5,3}(50)$} & \multicolumn{1}{c}{$\widehat{\alpha}_{BE}$}& \multicolumn{1}{c}{$\widehat{\beta}_{BE}$}& \multicolumn{1}{c}{$\widehat{R}_{5,3}(50)_{BE}$} \\
					\midrule
					I& 0.8965& 0.0230& 0.4309& 0.9051& 0.0235& 0.3944 \\
					II& 0.9907& 0.0176& 0.3647& 1.0015& 0.0169& 0.3629 \\
					III& 0.9402& 0.0205& 0.3946& 0.9491& 0.0199& 0.3916 \\
					\bottomrule
			\end{tabular}}
		\end{center}
		\vspace{-0.5cm}
	\end{table}
	
	\begin{table}
		\begin{center}
			\caption{The $95\%$ ACIs, HPD credible intervals for of $\alpha$, $\beta$ and $R_{5,3}(t)$ for various censoring schemes (CS).}
			\label{T6}
			\tabcolsep 7pt
			\small
			\scalebox{0.8}{
				\begin{tabular}{*{13}c*{11}{r@{}l}}
					\toprule
					\multicolumn{1}{c}{} &
					\multicolumn{2}{c}{$\alpha$} & \multicolumn{2}{c}{$\beta$} & \multicolumn{2}{c}{$R_{5,3}(50)$} \\
					\cmidrule(lr){2-3}\cmidrule(lr){4-5} \cmidrule(lr){6-7} 
					\multicolumn{1}{c}{CS} & \multicolumn{1}{c}{ACI} & \multicolumn{1}{c}{HPD}& \multicolumn{1}{c}{ACI} & \multicolumn{1}{c}{HPD}& \multicolumn{1}{c}{ ACI}& \multicolumn{1}{c}{HPD} \\
					\midrule
					I& (0.5276, 1.2654)& (0.7983, 0.9919)& (0.0018, 0.0480)& (0.0121, 0.0352)& (0.3653, 0.4966)& (0.3755, 0.4212) \\
					II& (0.6351, 1.3462)& (0.8866, 1.1222)& (0.0091, 0.0343)& (0.0101, 0.0238)& (0.2997, 0.4297)& (0.3368, 0.4051)  \\
					III& (0.6024, 1.2780)& (0.8389, 1.0639)& (0.0095, 0.0406)& (0.0139, 0.0258)& (0.3313, 0.4579)& (0.3581, 0.4117)\\
					\bottomrule
			\end{tabular}}
		\end{center}
		\vspace{-0.5cm}
	\end{table}
	
	\section{Conclusion}
	In this paper, the reliability analysis of the $K$-out-of-$N$: $G$ system composed of Weibull components under GPHC is considered. The MLEs and Bayes estimates
	of $\alpha$, $\beta$ and $R_{NK}(t)$ are obtained. The GPHC scheme not only allows to remove experimental units before failure but also guarantees certain number of failures in between a prefixed time. Thus, the efficiency of reliability
	inference can be improved within a pre-specified testing period. However, the computational complexity of parameter
	estimate and system reliability analysis have been increased due to GPHC scheme. To obtain MLEs of the parameters and the reliability function of the system, Newton-Raphson method has been employed to avoid complexity of calculations. Further, MCMC method has been used to obtain Bayes estimates and the associated HPD credible intervals. Then a Monte Carlo simulation study has been carried out to compare the performance of the estimates. From simulation study we can conclude that Bayes estimates perform better than MLEs in terms of MSEs and average length of the estimates. Finally a real data set has been analyzed for illustrative purposes. \\
	Using these discussions of $K$-out-of-$N$: $G$ system based on Weibull components, one can easily find the estimates for the system following exponential, Rayleigh and normal components by replacing the shape parameter as $1,2$, and $3.44$ respectively. In real life practice most of the data set has small number of samples, in these cases Weibull distribution is the best choice to handle. In future one can find the reliability of these system following other useful lifetime distributions. Further developments using the methods proposed in this paper can be applied to study the reliability analysis of $K$-out-of-$N$: $G$ balanced systems and shared-load $K$-out-of-$N$ system. 
	
	\section*{Acknowledgement}
	The authors thank the editor and two anonymous reviewers for their comments and valuable suggestions, which led to a considerable improvement in the content as well as the presentation of this manuscript. The author S. Dutta, thanks the Council of Scientific and Industrial Research (C.S.I.R.
	Grant No. 09/983(0038)/2019-EMR-I), India, for the financial assistantship received to carry out this research work. The authors thanks the research facilities received from the Department of Mathematics, National Institute of Technology Rourkela, India.
	
	\bibliography{myref1}

\begin{thebibliography}{}

\bibitem[\protect\citeauthoryear{Ashour, El-Sheikh, and Elshahhat}{Ashour
  et~al.}{2020}]{ashour2020inferences}
Ashour, S.~K., A.~A. El-Sheikh, and A.~Elshahhat (2020).
\newblock Inferences and optimal censoring schemes for progressively
  first-failure censored {N}adarajah-{H}aghighi distribution.
\newblock {\em Sankhya A\/}, 1--39.

\bibitem[\protect\citeauthoryear{Balakrishnan and Cramer}{Balakrishnan and
  Cramer}{2014}]{balakrishnan2014art}
Balakrishnan, N. and E.~Cramer (2014).
\newblock {\em The {A}rt of {P}rogressive {C}ensoring}.
\newblock Springer.

\bibitem[\protect\citeauthoryear{Balakrishnan and Kundu}{Balakrishnan and
  Kundu}{2013}]{balakrishnan2013hybrid}
Balakrishnan, N. and D.~Kundu (2013).
\newblock Hybrid censoring: {M}odels, inferential results and applications.
\newblock {\em Computational Statistics $\&$ Data Analysis\/}~{\em 57\/}(1),
  166--209.

\bibitem[\protect\citeauthoryear{Chen and Shao}{Chen and
  Shao}{1999}]{chen1999monte}
Chen, M.~H. and Q.~M. Shao (1999).
\newblock Monte {C}arlo estimation of {B}ayesian credible and {HPD} intervals.
\newblock {\em Journal of computational and Graphical Statistics\/}~{\em
  8\/}(1), 69--92.

\bibitem[\protect\citeauthoryear{Chen and Yang}{Chen and
  Yang}{2005}]{chen2005reliability}
Chen, Y. and Q.~Yang (2005).
\newblock Reliability of two-stage weighted-k-out-of-n systems with components
  in common.
\newblock {\em IEEE Transactions on Reliability\/}~{\em 54\/}(3), 431--440.

\bibitem[\protect\citeauthoryear{Childs, Chandrasekar, Balakrishnan, and
  Kundu}{Childs et~al.}{2003}]{childs2003exact}
Childs, A., B.~Chandrasekar, N.~Balakrishnan, and D.~Kundu (2003).
\newblock Exact likelihood inference based on type-{I} and type-{II} hybrid
  censored samples from the exponential distribution.
\newblock {\em Annals of the Institute of Statistical Mathematics\/}~{\em
  55\/}(2), 319--330.

\bibitem[\protect\citeauthoryear{Chiodo and Mazzanti}{Chiodo and
  Mazzanti}{2006}]{chiodo2006bayesian}
Chiodo, E. and G.~Mazzanti (2006).
\newblock Bayesian reliability estimation based on a {W}eibull stress-strength
  model for aged power system components subjected to voltage surges.
\newblock {\em IEEE Transactions on Dielectrics and Electrical
  Insulation\/}~{\em 13\/}(1), 146--159.

\bibitem[\protect\citeauthoryear{Cho, Sun, and Lee}{Cho
  et~al.}{2015}]{cho2015exact}
Cho, Y., H.~Sun, and K.~Lee (2015).
\newblock Exact likelihood inference for an exponential parameter under
  generalized progressive hybrid censoring scheme.
\newblock {\em Statistical Methodology\/}~{\em 23}, 18--34.

\bibitem[\protect\citeauthoryear{Cohen}{Cohen}{1963}]{cohen1963progressively}
Cohen, A.~C. (1963).
\newblock Progressively censored samples in life testing.
\newblock {\em Technometrics\/}~{\em 5\/}(3), 327--339.

\bibitem[\protect\citeauthoryear{Cui and Xie}{Cui and
  Xie}{2005}]{cui2005generalized}
Cui, L. and M.~Xie (2005).
\newblock On a generalized k-out-of-n system and its reliability.
\newblock {\em International Journal of Systems Science\/}~{\em 36\/}(5),
  267--274.

\bibitem[\protect\citeauthoryear{Dutta and Kayal}{Dutta and
  Kayal}{2022}]{dutta2022estimation}
Dutta, S. and S.~Kayal (2022).
\newblock Estimation of parameters of the logistic exponential distribution
  under progressive type-{I} hybrid censored sample.
\newblock {\em Quality Technology $\&$ Quantitative Management\/}~{\em
  19\/}(2), 234--258.

\bibitem[\protect\citeauthoryear{Elshahhat}{Elshahhat}{2017}]{elshahhat2017par%
ameters}
Elshahhat, A. (2017).
\newblock Parameters estimation for the exponentiated {W}eibull distribution
  based on generalized progressive hybrid censoring schemes.
\newblock {\em American Journal of Applied Mathematics and Statistics\/}~{\em
  5\/}(2), 33--48.

\bibitem[\protect\citeauthoryear{Epstein}{Epstein}{1954}]{epstein1954truncated}
Epstein, B. (1954).
\newblock Truncated life tests in the exponential case.
\newblock {\em The Annals of Mathematical Statistics\/}~{\em 25\/}(3),
  555--564.

\bibitem[\protect\citeauthoryear{Eryilmaz}{Eryilmaz}{2013}]{eryilmaz2013reliab%
ility}
Eryilmaz, S. (2013).
\newblock Reliability of a k-out-of-n system equipped with a single warm
  standby component.
\newblock {\em IEEE Transactions on Reliability\/}~{\em 62\/}(2), 499--503.

\bibitem[\protect\citeauthoryear{Franko, T{\"u}t{\"u}nc{\"u}, and
  Eryilmaz}{Franko et~al.}{2017}]{franko2017reliability}
Franko, C., G.~Y. T{\"u}t{\"u}nc{\"u}, and S.~Eryilmaz (2017).
\newblock Reliability of weighted k-out-of-n: {G} systems consisting of two
  types of components and a cold standby component.
\newblock {\em Communications in Statistics-Simulation and Computation\/}~{\em
  46\/}(5), 4067--4081.

\bibitem[\protect\citeauthoryear{G{\'o}rny and Cramer}{G{\'o}rny and
  Cramer}{2016}]{gorny2016exact}
G{\'o}rny, J. and E.~Cramer (2016).
\newblock Exact likelihood inference for exponential distributions under
  generalized progressive hybrid censoring schemes.
\newblock {\em Statistical Methodology\/}~{\em 29}, 70--94.

\bibitem[\protect\citeauthoryear{Hemmati and Khorram}{Hemmati and
  Khorram}{2013}]{hemmati2013statistical}
Hemmati, F. and E.~Khorram (2013).
\newblock Statistical analysis of the log-normal distribution under type-{II}
  progressive hybrid censoring schemes.
\newblock {\em Communications in Statistics-simulation and Computation\/}~{\em
  42\/}(1), 52--75.

\bibitem[\protect\citeauthoryear{Kotb}{Kotb}{2018}]{kotb2018bayesian}
Kotb, M.~S. (2018).
\newblock Bayesian prediction bounds for the exponential-type distribution
  based on generalized progressive hybrid censoring scheme.
\newblock {\em Stochastics and Quality Control\/}~{\em 33\/}(2), 93--101.

\bibitem[\protect\citeauthoryear{Kundu and Joarder}{Kundu and
  Joarder}{2006}]{kundu2006analysis}
Kundu, D. and A.~Joarder (2006).
\newblock Analysis of type-{II} progressively hybrid censored data.
\newblock {\em Computational Statistics $\&$ Data Analysis\/}~{\em 50\/}(10),
  2509--2528.

\bibitem[\protect\citeauthoryear{Li and Ni}{Li and
  Ni}{2008}]{li2008reliability}
Li, L. and J.~Ni (2008).
\newblock Reliability estimation based on operational data of manufacturing
  systems.
\newblock {\em Quality and Reliability Engineering International\/}~{\em
  24\/}(7), 843--854.

\bibitem[\protect\citeauthoryear{Lin, Chou, and Huang}{Lin
  et~al.}{2012}]{lin2012inference}
Lin, C.~T., C.~C. Chou, and Y.~L. Huang (2012).
\newblock Inference for the {W}eibull distribution with progressive hybrid
  censoring.
\newblock {\em Computational Statistics $\&$ Data Analysis\/}~{\em 56\/}(3),
  451--467.

\bibitem[\protect\citeauthoryear{Linhart and Zucchini}{Linhart and
  Zucchini}{1986}]{linhart1986model}
Linhart, H. and W.~Zucchini (1986).
\newblock {\em Model selection}.
\newblock John Wiley \& Sons.

\bibitem[\protect\citeauthoryear{Maswadah}{Maswadah}{2021}]{maswadah2021improv%
ed}
Maswadah, M. (2021).
\newblock Improved maximum likelihood estimation of the shape-scale family
  based on the generalized progressive hybrid censoring scheme.
\newblock {\em Journal of Applied Statistics\/}, 1--20.

\bibitem[\protect\citeauthoryear{Mohamed, Abu-Youssef, Ali, and Abd
  El-Raheem}{Mohamed et~al.}{2018}]{mohamed2018inference}
Mohamed, A. E.-R., S.~Abu-Youssef, N.~S. Ali, and A.~Abd El-Raheem (2018).
\newblock Inference on constant-stress accelerated life testing based on
  geometric process for extension of the exponential distribution under
  type-{II} progressive censoring.
\newblock {\em Pakistan Journal of Statistics and Operation Research\/}~{\em
  14\/}(2), 233--251.

\bibitem[\protect\citeauthoryear{Qihong and Junrong}{Qihong and
  Junrong}{2014}]{qihong2014parameter}
Qihong, D. and L.~Junrong (2014).
\newblock Parameter estimation for a k-out-of-n: F system.
\newblock {\em Communications in Statistics-Simulation and Computation\/}~{\em
  43\/}(1), 99--114.

\bibitem[\protect\citeauthoryear{Roy and Gupta}{Roy and
  Gupta}{2021}]{roy2021reliability}
Roy, A. and N.~Gupta (2021).
\newblock Reliability function of k-out-of-n system equipped with two cold
  standby components.
\newblock {\em Communications in Statistics-Theory and Methods\/}~{\em
  50\/}(24), 5759--5778.

\bibitem[\protect\citeauthoryear{Singh, Singh, and Yadav}{Singh
  et~al.}{2015}]{singh2015reliability}
Singh, S.~K., U.~Singh, and A.~S. Yadav (2015).
\newblock Reliability estimation and prediction for extension of exponential
  distribution using informative and non-informative priors.
\newblock {\em International Journal of System Assurance Engineering and
  Management\/}~{\em 6\/}(4), 466--478.

\bibitem[\protect\citeauthoryear{Tomer and Panwar}{Tomer and
  Panwar}{2015}]{tomer2015estimation}
Tomer, S.~K. and M.~Panwar (2015).
\newblock Estimation procedures for {M}axwell distribution under type-{I}
  progressive hybrid censoring scheme.
\newblock {\em Journal of Statistical Computation and Simulation\/}~{\em
  85\/}(2), 339--356.

\bibitem[\protect\citeauthoryear{Wang}{Wang}{2016}]{wang2016conditional}
Wang, Y. (2016).
\newblock Conditional k-out-of-n systems with a cold standby component.
\newblock {\em Communications in Statistics-Theory and Methods\/}~{\em
  45\/}(21), 6253--6262.

\bibitem[\protect\citeauthoryear{Zhao, Xiang, Cheng, and Guo}{Zhao
  et~al.}{2019}]{zhao2019estimation}
Zhao, Q., J.~Xiang, Z.~Cheng, and B.~Guo (2019).
\newblock Estimation of lifetime and residual life of typical system with
  {W}eibull distributed components.
\newblock {\em Systems Engineering and Electronics\/}~{\em 41\/}(7), 1665--71.

\bibitem[\protect\citeauthoryear{Zhao and Cui}{Zhao and
  Cui}{2010}]{zhao2010reliability}
Zhao, X. and L.~Cui (2010).
\newblock Reliability evaluation of generalised multi-state k-out-of-n systems
  based on {FMCI} approach.
\newblock {\em International Journal of Systems Science\/}~{\em 41\/}(12),
  1437--1443.

\end{thebibliography}
\end{document}